# Dropout MPC: An Ensemble Neural MPC Approach for Systems with Learned Dynamics

Spyridon Syntakas and Kostas Vlachos, *Member, IEEE*

*Abstract*— Neural networks are lately more and more often being used in the context of data-driven control, as an approximate model of the true system dynamics. Model Predictive Control (MPC) adopts this practise leading to neural MPC strategies. This raises a question of whether the trained neural network has converged and generalized in a way that the learned model encapsulates an accurate approximation of the true dynamic model of the system, thus making it a reliable choice for model-based control, especially for disturbed and uncertain systems. To tackle that, we propose Dropout MPC, a novel sampling-based ensemble neural MPC algorithm that employs the Monte-Carlo dropout technique on the learned system model. The closed loop is based on an ensemble of predictive controllers, that are used simultaneously at each time-step for trajectory optimization. Each member of the ensemble influences the control input, based on a weighted voting scheme, thus by employing different realizations of the learned system dynamics, neural control becomes more reliable by design. An additional strength of the method is that it offers by design a way to estimate future uncertainty, leading to cautious control. While the method aims in general at uncertain systems with complex dynamics, where models derived from first principles are hard to infer, to showcase the application we utilize data gathered in the laboratory from a real mobile manipulator and employ the proposed algorithm for the navigation of the robot in simulation.

## I. INTRODUCTION

Even before the rise of deep learning, neural networks have been successfully employed in the context of MPC [1], [2]. More so, nowadays, the efficiency of modern toolkits for non-linear optimization, specialized hardware and the advances of deep learning have made neural MPC more trivially applicable even to real time robot control problems [3]. However the success of MPC, as a model-based control strategy, is subject to the accuracy of the model used.

Although system models approximated with neural networks can be in many cases a lot more accurate compared to a first principle derived model, there are many occasions that even these deep learning approaches can fail to capture accurately the system model due to poor generalization. These cases can occur for example due to lack of data from the real plant, especially in the case of disturbed systems, where a sufficient number of data-points has to be gathered. This is a particularly often case in many robotic applications, where efficient data gathering from real world applications is not even possible due to safety critical issues. In addition to that, closed-loop data gathering, despite being safer, can lead to biased models that do not generalize well [2]. Modern simulators try to tackle the problem of efficient data gathering, but sim-to-real gap still remains an open problem [4]. Limited data also can lead to overfitting, especially in the case of deep architectures, which can become problematic for many applications.

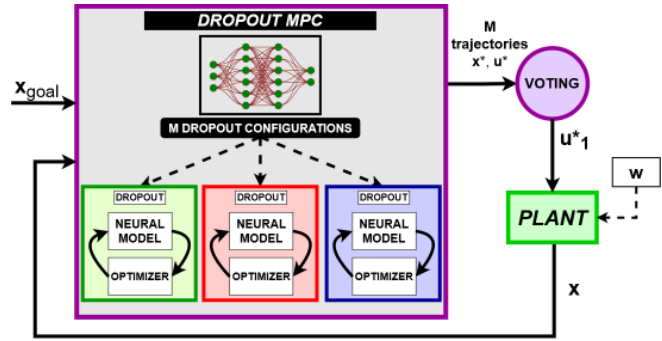

Fig. 1: Schematic representation of the Dropout MPC method.

A way to avoid some of these issues, especially overfitting and efficient data utilization, is to employ ensembles and ensemble learning, i.e., to train weak learners that as a whole form a strong learner that acts as a better predictive model with reduced variance (bagging), bias (boosting) and improved predictions (stacking) [5]. Ensembles bring improvements to a variety of classification and regression problems in many different fields of application [6], especially nowadays that deep learning is the predominant learning approach [7]. In this work, instead of training separate weak learners, a single neural network is trained and employing the Monte-Carlo dropout technique an ensemble of predictive controllers is formed, by sampling from the parametric posterior and utilizing different model realizations as predictive models.

More specifically, we propose a novel efficient neural MPC strategy that employs the Monte-Carlo dropout technique to form an ensemble of neural MPC laws at each time-step. Each one of these neural controllers provides an optimized trajectory by employing a different realization of the learned model via posterior sampling. This predictive ensemble, votes at each time-step in order define the control input to the system based on a weighted average, sharing

*We acknowledge support of this work by the project "Dioni: Computing Infrastructure for Big-Data Processing and Analysis." (MIS No. 5047222) which is implemented under the Action "Reinforcement of the Research and Innovation Infrastructure", funded by the Operational Programme "Competitiveness, Entrepreneurship and Innovation" (NSRF 2014-2020) and co-financed by Greece and the European Union (European Regional Development Fund).

Spyridon Syntakas and Kostas Vlachos are with the Department of Computer Science and Engineering, University of Ioannina, 45110 Ioannina, Greece. (Email: {ssyntakas; kostaswl}@cse.uoi.gr)

intuition with the vanilla MPPI algorithm [8]. The proposed method grands advantages over the traditional neural MPC by design:

- By employing the MC-dropout technique and using an ensemble of MPC laws based on posterior sampling, the proposed method counterbalances the negative effects of a single over-fitted model.
- Weighted voting between different realizations of the optimized trajectory, leads to control inputs that are more reliable compared to committing to a single over-fitted model. The ensemble that votes the control input encapsulates knowledge in a more expressive way than that of a single network, due to posterior sampling.
- The proposed algorithm offers a way to estimate future uncertainty. This is inherent from the MC-dropout technique and a careful quantification of the uncertainty can lead to cautious control in safety-critical applications.

In addition, as the ensemble is formed using different realizations of the learned model, only one neural network needs to be trained, which simplifies a lot the learning process compered to traditional ensemble forming techniques.

As estimation of both the epistemic [9] and aleatory uncertainty in MPC is a topic of interest, the MC-dropout method has been employed in some works. [10] focuses on uncertainty estimation on imitation learning of MPC demonstrations for insulin control via MC-dropout. In [11] the MC-dropout technique is used during visual navigation where the output variance of a network is used for the assessment of unsafe conditions. The intuition behind the employment of MC-dropout in these works differs from our presented approach. A similar employment of the MC-dropout technique to that of the presented work is done in [12], where a nominal MPC based on a first principles derived model is used for control and the MC-dropout technique is used to quantify the uncertainty of the nominal horizon. Compared to the presented method where the robot is controlled by a neural ensemble of MPC laws, in [12] the robot is controlled by a nominal MPC and there is no notion of ensemble voting for the control input at each time-step.

## II. THEORETIC BACKGROUND

### A. Neural MPC

The systems of interest have the following general form

$$\dot{\mathbf{x}} = f(\mathbf{x}, \mathbf{u}) + \mathbf{w} \quad (1)$$

i.e, a non-linear disturbed system with $\mathbf{x}$ denoting the state vector, $\mathbf{u}$ the control input vector and $\mathbf{w}$ representing disturbances. The standard MPC paradigm uses an $f(.)$ derived from first principles, but the neural MPC paradigm replaces the first principle derived model with a neural network in order to capture complex nonlinear system dynamics. Thus, given a goal state of the system denoted as $\mathbf{x}_{ref}$, the neural MPC law is defined as:

$$\min_{\mathbf{x},\mathbf{u}} V_f(\mathbf{x}_{t+N|t} - \mathbf{x}_{ref}) + \sum_{k=0}^{N-1} l(\mathbf{x}_{t+k|t} - \mathbf{x}_{ref}, \mathbf{u}_{t+k|t}) \quad (2)$$

$$\mathbf{x}_{t+k+1|t} = f_{NN}(\mathbf{x}_{t+k|t}, \mathbf{u}_{t+k|t}) \quad (3)$$

$$\mathbf{x}_{t+k|t} \in \mathcal{X} \quad (4)$$

$$\mathbf{u}_{t+k|t} \in \mathcal{U} \quad (5)$$

$$\mathbf{x}_{t|t} = \mathbf{x}_t \quad (6)$$

$$\mathbf{x}_{t+N|t} = \mathbf{x}_{ref} \quad (7)$$

where $f_{NN}(.)$ is the neural network approximation of the true system dynamics, i.e., the learned model that is employed as the prediction model. $\mathbf{x}_{t+k|t}$ denotes the state vector at time $t+k$, predicted at time $t$, if $\mathbf{u}_{t:t+k-1|t}$ is applied to the system model from the current state $\mathbf{x}_t$ for $k = \{0, 1, 2, .., N-1\}$, with $N$ being the control horizon. Solving (2) at time $t$, provides the optimal finite horizon control trajectory $\mathbf{u}^*_{t:t+N-1} = (\mathbf{u}^*_{t|t}, \mathbf{u}^*_{t+1|t}, ..., \mathbf{u}^*_{t+N-1|t})$. The first input is applied to the system leading to the receding horizon control law $\mathbf{u}(t) = \mathbf{u}^*_{t|t}(\mathbf{x}_t)$. The nonlinear program is solved at each time-step employing the neural network as the prediction model as described in eq. (3). The cost to be minimized is designed as in expression (2) and is a convex quadratic function. $l(\mathbf{x}_{t+k|t} - \mathbf{x}_{des}, \mathbf{u}_{t+k|t})$ denotes the stage cost and has the form of $(\mathbf{x} - \mathbf{x}_{des})^T Q (\mathbf{x} - \mathbf{x}_{des}) + \mathbf{u}^T R \mathbf{u}$, with $Q$ and $R$ being diagonal and positive definite weighting matrices. $V_f(\mathbf{x}_{t+N|t} - \mathbf{x}_{des})$ represents the terminal cost which together with the terminal equality constraint (7) are the "stability axioms" [13]. The terminal cost is calculated using the Discrete Algebraic Riccati Equation (DARE) of the linearized system at the equilibrium. Linear approximations of neural networks can provide valuable information about the networks behaviour [14], [15] and they are employed in this work for stability guaranties. $\mathcal{X}$ and $\mathcal{U}$ denote compact constraint sets containing the equilibrium, i.e., $\mathbf{x}_{ref} \in \mathcal{X} \times \mathcal{U}$.

### B. Monte-Carlo dropout theory

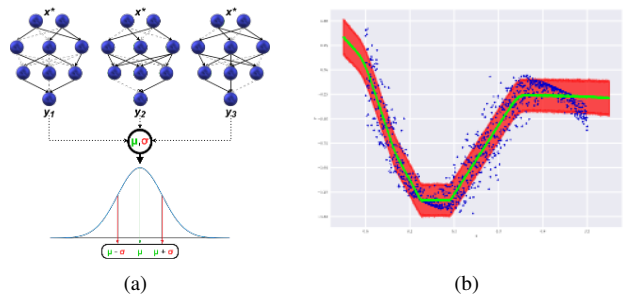

(a)          (b)

Fig. 2: Schematic representation of the MC-dropout inference

An approach to achieve inference in a way that also estimates uncertainty in a Bayesian manner, is to infer a predictive distribution of the form $p(\mathbf{y}|\mathbf{x}, \mathbf{D})$, with $\mathbf{y}$ denoting the target variable, $\mathbf{x}$ the input to the neural network and $\mathbf{D}$ denoting the $N$ training examples $(\mathbf{x}_i, \mathbf{y}_i)$ with $i = 1, 2, 3, .., N$. A way to learn this predictive distribution is by learning the parametric posterior distribution $p(\theta|\mathbf{D})$ i.e., a distribution over the neural network parameters. Monte-Carlo dropout method [16] offers an intuitive way to approximate the parametric posterior as $q(\theta|\mathbf{D})$. Dropout is a regularization technique applied to neural networks by randomly switching off neurons based on Bernoulli trials, during training. Using dropout before each neural network layer we obtain neural network realizations that correspond to samples $\theta_j$ from the approximate posterior distribution $q(\theta|\mathbf{D})$. Thus, using sampling, the true posterior that is described as:

$$p(\mathbf{y}|\mathbf{x}) = \int p(\mathbf{y}|\mathbf{x}, \mathbf{D}) q(\theta|\mathbf{D})\, d\theta \tag{8}$$

can be approximated using Monte Carlo as,

$$p(\mathbf{y}|\mathbf{x}) = \frac{1}{M} \sum_{j=1}^{M} p(\mathbf{y}|\mathbf{x}, \theta_j), \theta_j \sim q(\theta|\mathbf{D}) \tag{9}$$

In practice, inference is done by performing M stochastic forward passes through the network for the same input, using M different dropout configurations, and averaging the results. By assuming Gaussian distributed likelihood $p(\mathbf{y}|\mathbf{x}, \mathbf{D})$, a mean and a standard deviation $(\mu, \sigma)$ of the output $\mathbf{y}$ can be obtained for a given input $\mathbf{x}$, as depicted in Fig. 2.

In this work, we are not only interested in the uncertainty estimation capabilities of the MC-dropout technique as a Bayesian approximation. In a similar manner to the MC-dropout posterior sampling, the presented method forms the ensemble of prediction models that are used in the receding horizon control law, employing during trajectory optimization each one of the M different model realizations, as those depicted in Fig. 2(a).

## III. DROPOUT MPC

A schematic representation of the method is depicted in Fig. 1 and an algorithmic overview can be seen in Algorithm 1. A neural network is trained offline to learn the model of the system employing data from the plant. Although, state estimation should be used in a real application, in this section full state feedback is assumed for simplicity.

### A. Modeling from data

A Dense Neural Network (DNN) architecture is used to learn the true model of the system. In order to train the neural network, data are gathered from random walks as well as trajectories obtained using teleoperation of the robotic system. Thus a dataset $D$ of the form $D = \{\mathbf{x}_t, \mathbf{u}_t, \dot{\mathbf{x}}\}$ is created, with $\dot{\mathbf{x}}$ being the target variable, given full state observability. $\dot{\mathbf{x}}$ can be obtained as the difference of two sequential states quotient with a small time-step $T$. If data from simulations are to be used for training, extra care should be taken, in order to gather a dataset that is not biased and the data distribution matches the real world application. For that reason, data from simulation and data gathering based on nominal model closed-loop trajectories is avoided in this work.

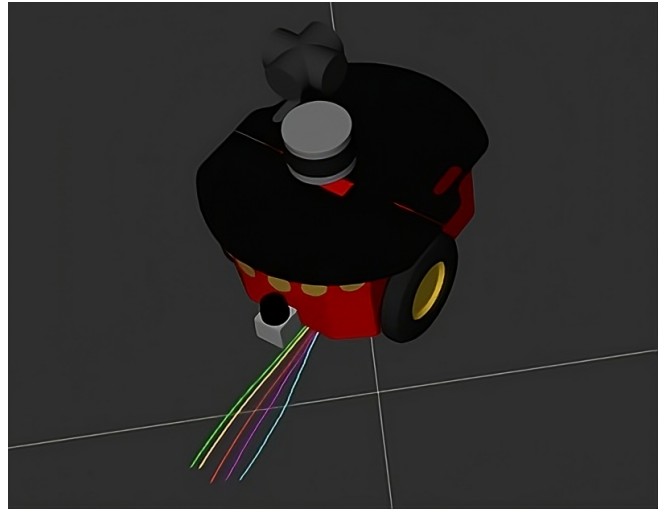

Fig. 3: State horizons at a given time-step produced by the ensemble during navigation of a simulated Pioneer P3-DX.

### B. Ensemble formation & Voting for the control input

Using the trained network and employing the MC-dropout method, M different realizations of the system model are obtained. At each time-step, each one of these M different system model realizations are used simultaneously, as predictive models, in a neural MPC manner, and the ensemble is created. In total M optimized control trajectories of the form $\mathbf{u}^*_{t:t+N-1} = (\mathbf{u}^*_{t|t}, \mathbf{u}^*_{t+1|t}, ..., \mathbf{u}^*_{t+N-1|t})$ are inferred at each time-step. The first element, $\mathbf{u}^*_{t|t}$, of each one these M optimized trajectories is used in the voting scheme. The voting is conducted using a weighted average, with the weights being proportional to the M horizon costs. An example of such state horizons derived at a specific time-step during the navigation of a Pioneer 3-DX mobile robot in the Gazebo simulator is depicted in Fig. 3

More specifically, the M horizon costs, that are employed to calculate the weights, are standardized to zero mean and unit standard deviation by subtracting the mean from each value and divide by the standard deviation. Then, the weight of each control input proposed by each member of the ensemble is calculated as

$$w_i = e^{-\frac{1}{\kappa} cost_i} \tag{10}$$

with $i = \{0, 1, 2, ..., M\}$ and $\kappa \in \mathbb{N}$. The weights are normalized using Min-Max scaling and the control input proposed by the ensemble at each time-step is calculate as:

$$\mathbf{u}_{ensemble} = \frac{\sum_{i=0}^{M} w_i \mathbf{u}_i}{\sum_{i=0}^{M} w_i} \tag{11}$$

The final control input to the system is calculated as a weighted summation of the control input inferred using the full DNN and the control input proposed by the ensemble as:

$$\mathbf{u}_{final} = \lambda \mathbf{u}_{NN} + \xi \mathbf{u}_{ensemble} \tag{12}$$

with $\lambda > \xi$ and $\mathbf{u}_{NN}$ denoting the standard neural MPC control input at each time-step.

*C. Uncertainty estimation of the next state*

Using the M different realizations of the neural model and employing the inferred control input using the Runge–Kutta method (RK4) from the current state, $\mathbf{x}_{t|t}$, M different realizations of the possible next state $\mathbf{x}^i_{t|t+1}$ are obtained, with $i \in [1, 2, 3, ..., M]$. Assuming a bivariate normal distribution, a way to quantify uncertainty of the next state is by calculating its standard deviation. Based on that the predictive algorithms can be enhanced with a meaningful quantification of the next state uncertainty.

**Algorithm 1:** Dropout MPC

**Input** : Goal state $\mathbf{x_{ref}}$
**Input** : Current state $\mathbf{x_{t|t}}$
**Input** : Neural Model $f_{NN}$
**Input** : Cost Function $l$
**Input** : Constraints $g(\mathbf{x}, \mathbf{u})$
**Input** : Accuracy threshold $\epsilon$
**Output:** Voted control input $\mathbf{u_{final}}$
**Output:** Next state uncertainty $\mathbf{x}^\sigma$

1 **while** $\|\mathbf{x_{t|t}} - \mathbf{x_{ref}}\| \leq \epsilon$ , *at each time-step t* **do**
    /* Solve neural MPC – No dropout */
2    $\mathbf{x^{**}_{t:t+N}}, \mathbf{u^{**}_{t:t+N-1}} \leftarrow \mathbf{MPC}(\mathbf{x_{t|t}}, f_{NN}, g(.))$
    /* Ensemble formation */
3    **for** *i in range M* **do**
      /* Get a different model realization via dropout */
4      $f^i_{NN} \leftarrow \mathbf{ApplyDropout}(f_{NN})$
      /* Solve neural MPC */
5      $\mathbf{x^{*,i}_{t:t+N}}, \mathbf{u^{*,i}_{t:t+N-1}} \leftarrow \mathbf{MPC}(\mathbf{x_{t|t}}, f^i_{NN}, g(.))$
      /* Calculate corresponding weight */
6      $w_i \leftarrow \mathbf{CalculateWeight}(l^i)$
7    **end**
    /* Calculate the ensemble control input via voting */
8    $\mathbf{u_{ensemble}} \leftarrow \mathbf{WeightedAvg}(w_{i:M}, u^{*,i:M}_{t|t})$
    /* Get final input */
9    $\mathbf{u_{final}} \leftarrow \lambda \mathbf{u^{**}_{t|t}} + \xi \mathbf{u_{ensemble}}$
    /* Calculate next state uncertainty */
10    $\mathbf{x}^\sigma \leftarrow \mathbf{Std}(\mathbf{x^{*,i:M}_{t|t+1}})$
11 **end**

## IV. EXPERIMENTAL RESULTS

As a proof-of-concept, the proposed method is applied for the navigation and parallel parking of a mobile manipulator in simulation with data gathered from the real mobile manipulator depicted in Fig. 4. More specifically, the mobile manipulator consists of a Jackal mobile base from Clearpath equipped with a Gen3 lite 6-Dof manipulator by Kinova and various sensors. As we are interested in the navigation of the robot, we focus only on the degrees of freedom of the mobile base, thus only the odometry and IMU sensory measurements are employed. ROS Noetic [17] is used throughout the implementation of the experiments, from data gathering to software development and deployment of the method.

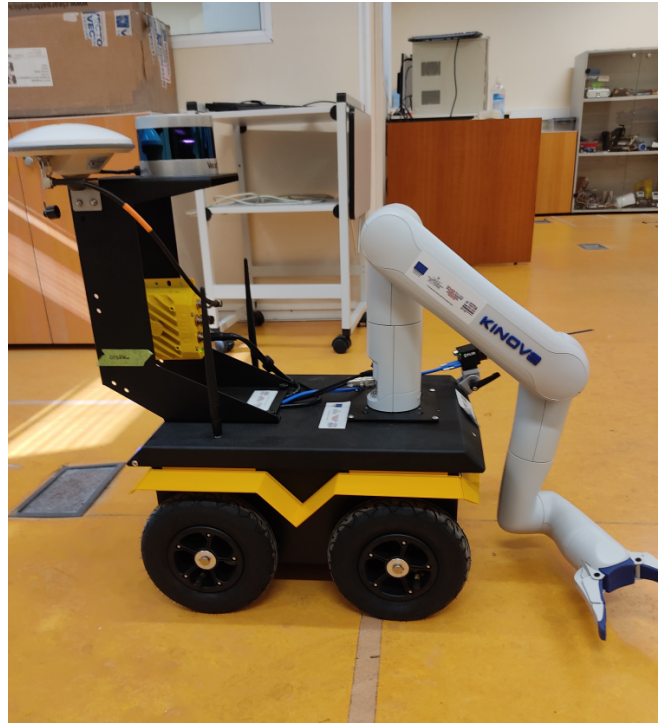

Fig. 4: Jackal with Gen3 lite - The mobile manipulator used for data gathering.

Although the effectiveness of the method is demonstrated for the task of controlling the mobile base, where data gathering is not a difficult task in general, the benefit of the method stems from the fact that ensemble modeling has the ability to perform well in cases where data are limited.

*A. Data Gathering and training*

The data what are used for training, are obtained by teleoperating the mobile base using velocity commands in the laboratory. The pose, denoted as $x, y, \theta$ together with the linear and angular velocity commands, denoted as $v, \omega$ were recorded employing rosbag files. The odometry and IMU sensors were fused using an extended Kalman filter (EKF) to get more accurate measurements of the robot's pose. Using the frequency of the sensory measurements that is $h = 53Hz$ and the pose difference quotient with $1/h$, a data set is obtained of the form $D = \{x, y, \theta\ v, \omega, \dot{x}, \dot{y}, \dot{\theta}\}$. Furthermore, some additional data are obtained in the same manner but by performing random walks of the mobile base. Data obtained by random walks are used to reduce teleoperation expert bias. The final dataset consists of 43628 examples. Data gathering from closed loop autonomous operation of the robot is avoided as the controller bias can lead to data not being representative of the system, thus generalization can be crippled [2].

Pytorch [18] is the framework of choice regarding learning and modeling. The network of choice is as shallow as possible and specifically a network with 1-hidden layer consisting of 30 nodes is used. This choice is supported by the fact that the algorithm should not depend on specialized hardware to run in real time and preferably run only on a CPU, as many small autonomous systems do not have a dedicated GPU. Sigmoid is used as the non-linearity. The choice of the sigmoid function as the non-linearity, opposed to the more commonly nowadays used ReLU, is done as it makes the solution of the non-linear program at each time-step more lightweight computationally. That leads to higher control frequency. If a system is equipped with hardware that can accelerate inference, deeper networks may be employed. The dataset was split to 80-20 training-testing ratio and the model was trained for 600 epochs using Adam as the optimizer and standard Mean Square Error as the loss function.

### B. Design of the Drop-out MPC

An ensemble of $M = 10$ predictive controllers is employed using 10 different configurations of the neural networks based on the MC-dropout as described in section III. The MPCs that form the ensemble are all implemented using CasADi [19], an open-source tool for nonlinear optimization and algorithmic differentiation. Multiple shooting is employed as the transcription method for the optimal control problem and Euler's forward method is used for numerical integration. In general, the ensemble can consist of more members if the computational power is available but we choose $M = 10$ as the algorithm runs on a single Ryzen 5 5600G CPU though all the experiments. This does not cripple the uncertainty estimation capabilities of the method. The drop-out rate is chosen to be equal to 0.2 and is conducted based on Bernoulli trails at each time-step. As for the weighting of the control input in (12), we choose $\lambda = 0.7$ and $\xi = 0.3$. A terminal cost can be used as a "stability axiom" based on linear approximation of the neural network [14], [15].

### C. Point-goal navigation

As a first proof of concept, the method is applied for point-goal navigation of the mobile manipulator. The mobile base initial pose is $[0m, 0m, 0rads]$ and the target pose is $[1m, 2m, \frac{\pi}{4}rad]$. Figs. 5, 6, 8 depict a comparison between a standard neural MPC and the Drop-out MPC technique for that same task. As it can be seen the two methods have similar performance but the Drop-out method comes with the additional aspect of in build uncertainty estimation as depicted in Fig. 7.

### D. Parallel parking

A second proof of concept is conducted by applying the method for the parallel parking of the mobile manipulator. The mobile base initial pose is $[0m, 0m, 0rads]$ and the target pose is $[0m, 1m, 0rad]$. Figs. 9, 10, 12 depict a comparison between a standard neural MPC and the Drop-out MPC technique for the same a task. The two tasks achieve similar performance but Drop-out MPC comes again with the advantage of uncertainty estimation.

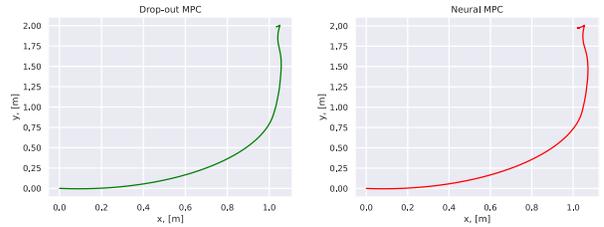

Fig. 5: Point-goal navigation: Paths comparison.

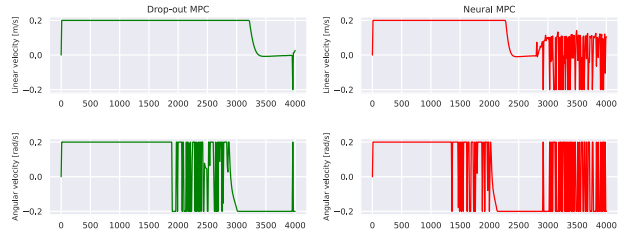

Fig. 6: Point-goal navigation: Command inputs.

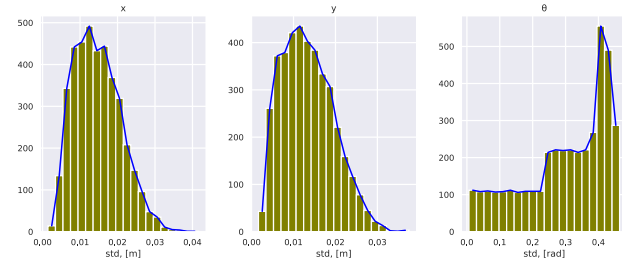

Fig. 7: Point-goal navigation: State uncertainty.

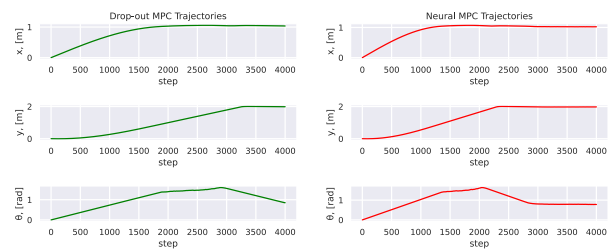

Fig. 8: Point-goal navigation: State trajectories.

### V. CONCLUSION

The paper presents the design of the Dropout MPC method, an ensemble sampling-based neural MPC strategy. The method combines intuition from predictive control methodologies, the Monte-Carlo dropout technique and ensemble learning and builds on their advantages. Thus, it offers by design more reliable predictive control due to the ensemble voting and comes with in build uncertainty estimation capabilities while maintaining the stability guarantees of traditional MPC. In addition, it is general and can be trivially applied to any system with learned dynamics that employ neural networks. As a test case, the method is applied for the navigation of a mobile manipulator. It accomplishes, similar behavior to that of classic neural MPC but with the additional benefit of uncertainty quantification that leads to more cautious control as well as avoiding the negative effects of an over-fitted neural network as the predictive model.

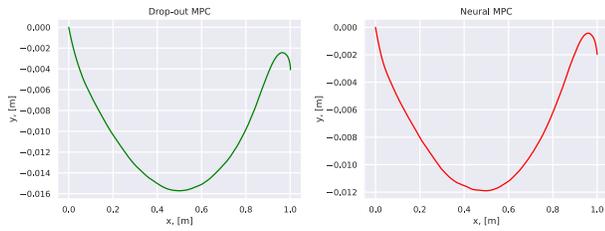

Fig. 9: Parallel parking: Paths comparison.

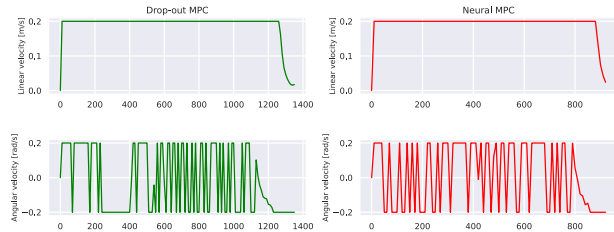

Fig. 10: Parallel parking: Command inputs.

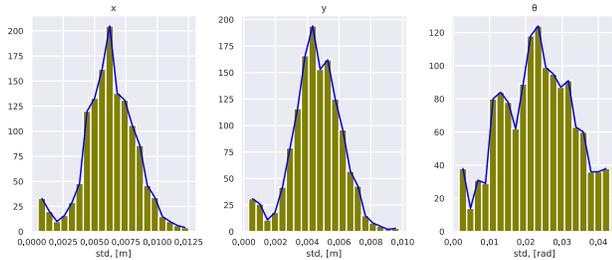

Fig. 11: State uncertainty.

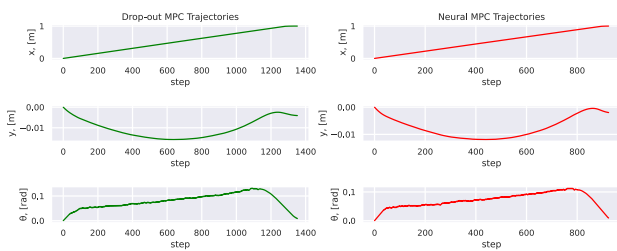

Fig. 12: Parallel parking: State trajectories.